\documentclass{pasj01}%[dvipdfmx]
\draft 
\Received{$\langle$2021 November 16$\rangle$}
\Accepted{$\langle$2021 February 4$\rangle$}
\Published{$\langle$publication date$\rangle$}
\usepackage{lscape}

\newcommand{\tabref}[1]{Table~\ref{#1}}
\newcommand{\equref}[1]{Eq~(\ref{#1})}
\newcommand{\figref}[1]{Figure~\ref{#1}}

\usepackage[hang,small,bf]{caption}
\usepackage[subrefformat=parens]{subcaption}
\begin{document}

\title{Measurements of Chromospheric Mg\,\emissiontype{I} Emission Lines of Zero-Age Main-Sequence Stars}
\author{Mai Yamashita${}^1$, Yoichi Itoh${}^1$}%, ${}^2$Yuhei Takagi
\altaffiltext{}{${}^1$Nishi-Harima Astronomical Observatory, Center for Astronomy, University of Hyogo, 407-2 Nishigaichi, Sayo, Sayo, Hyogo 679-5313 }
%\altaffiltext{}{${}^2$Subaru Telescope, National Astoronomical Observatory of Japan, 650 North A'ohoku Place, Hilo, HI 96720, U.S.A.}
\email{yamashita@nhao.jp}

%stars: pre-main sequence --- 

\maketitle

\begin{abstract}
The chromosphere is the active atmosphere in which energetic eruption events, such as flares, occur. Chromospheric activity is driven by the magnetic field generated by stellar rotation and convection. The relationship between chromospheric activity and the Rossby number, the ratio of the rotational period to the convective turnover time, has been extensively examined for many types of stars, by using narrow chromospheric emission lines, such as the Ca\, \emissiontype{II} lines and the Mg\, \emissiontype{II} h and k lines. However, the stars with small Rossby numbers, i.e., stars with rapid rotations and/or long convective turnover times, show constant strengths of such lines against the Rossby number. In this study, we investigate the infrared Mg\, \emissiontype{I} emission lines at $8807 \, \mathrm{\AA}$ of 47 zero-age main-sequence (ZAMS) stars in IC 2391 and IC 2602 using the archive data of the Anglo-Australian Telescope at the University College London Echelle Spectrograph. After subtracting the photospheric absorption component, the Mg\, \emissiontype{I} line is detected as an emission line for 45 ZAMS stars, whose equivalent widths are between $0.02\, \mathrm{\AA}$ and $0.52\, \mathrm{\AA}$. A total of 42 ZAMS stars show the narrower Mg\,\emissiontype{I} emission lines instead of the Ca\, \emissiontype{II} infrared triplet emission lines, suggesting that they are formed at different depths. The ZAMS stars with smaller Rossby numbers show stronger Mg\, \emissiontype{I} emission lines. The Mg\, \emissiontype{I} emission line is not saturated even in ``the saturated regime of the Ca\, \emissiontype{II} emission lines,''  i.e., Rossby number $< 10^{-1.1}$. The Mg\, \emissiontype{I} emission line is considered to be a good indicator of chromospheric activity, particularly for active objects. 
\end{abstract}

%\linenumbers

% 以下で本文を別ファイルから挿入する
\section{Introduction} 
\label{intro}

The chromosphere is the active atmosphere in which flares and other energetic eruption phenomena occur. It is claimed that chromospheric activity is driven by the magnetic fields induced by the dynamo process. In the dynamo process, the Coriolis force (= rotational moment $\times$ convection velocity) balances the Lorentz force (= current $\times$ magnetic strength / density of plasma) \citep{ba96}. Stellar rotation and convection are considered to be the main processes that drive the evolution of magnetic activities. \citet{noyes} used the Rossby number, $N_{\rm R}$, as an indicator of stellar activity. It is defined as $P/\tau_{\rm c}$, where $P$ is the stellar rotational period and $\tau_{\rm c}$ is the convective turnover time. $N_{\rm R}$ can be approximated as the inverse square of the dynamo number, $N_D$, the wave solution of the dynamo equation. Magnetic fields develop when $|N_D| > 1$. 

The relationship between chromospheric line strength and the Rossby number has been extensively examined for main-sequence stars. \citet{s72} found that the luminosity of the Ca\,\emissiontype{II} emission lines and rotational velocity of a solar-type star with an age of $10^8 \, \mathrm{yr}$ are one order of magnitude larger than those of stars with an age of $10^{10} \, \mathrm{yr}$. \citet{l79} obtained profiles of the Ca\,\emissiontype{II} infrared triplet (IRT) line at $\lambda 8542\, \mathrm{\AA}$ in 49 main-sequence or giant stars of spectra type F9-K3. For active chromospheric stars, they found that the line cores are filled in compared to quiet chromospheric stars of the same spectral type. They claimed that a better way of testing for chromospheric emission on the photospheric absorptinon is to subtract the profiles of similar spectral type stars. As a result, they showed strong evidence of chromospheric emissoin component beyond the line core for active stars. \citet{so93} revealed strong Ca\,\emissiontype{II} IRT ($\lambda 8498, \, 8542, \, 8662 \, \mathrm{\AA}$) emission lines of low-mass stars in a young open cluster M45. \citet{m09} also detected Ca\,\emissiontype{II} IRT emission lines of low-mass stars in young open clusters IC 2391 and IC 2602. The cluster members are considered to be on the zero-age main-sequence (ZAMS) or in the last evolution phase to the ZAMS. \citet{so93} and \citet{m09} calculated $R^{\prime}_{\rm IRT}$ from the equivalent widths (EQWs). $R^{\prime}_{\rm IRT}$ describes the ratio of the surface flux of the Ca\,\emissiontype{II} IRT emission lines to the stellar bolometric luminosity. They found that for stars with $N_{\rm R} \geq 10^{-1.1}$, $R^{\prime}_{\rm IRT}$ decreases with increasing $N_{\rm R}$. This region is called the unsaturated regime. In contrast, $R^{\prime}_{\rm IRT}$ is constant at levels of approximately $10^{-4.2}$ for stars with $N_{\rm R} \leq 10^{-1.1}$. This region is called the saturated regime. \citet{m09} suggested that the chromosphere is completely filled by the emitting regions for the stars in the saturated regime. %\footnote{\citet{l79}では晩期型星の Ca\,\emissiontype{II} ($\lambda 8542\, \mathrm{\AA}$) emission lines の吸収線を解析した。そしてスペクトル同士の減算は、彩層の活動度の比較として良い方法であると主張した。晩期型星のCa\,\emissiontype{II}は吸収線を示したが、類似したスペクトル型とluminosity classを持つ晩期型星同士でスペクトルを引くと彩層輝線が出現した。(またスペクトル同士を割るよりも減算のほうがbetter wayであると述べた。)}

\citet{y20} investigated the relation between $N_{\rm R}$ and the Ca\,\emissiontype{II} IRT emission lines of $60$ pre-main sequence (PMS) stars. Only three PMS stars showed broad and strong emissions, indicative of large mass accretion. Most of the PMS stars present narrow and weak emissions, suggesting that their emission lines are formed in the chromosphere. All their Ca\,\emissiontype{II} IRT emission lines have $R^{\prime}_{\rm IRT} \sim 10^{-4.2}$, which is as large as the maximum $R^{\prime}_{\rm IRT}$ of ZAMS stars. The PMS stars show $N_{\rm R} < 10^{-0.8}$ and constant $R^{\prime}_{\rm IRT}$ against $N_{\rm R}$, i.e., their Ca\,\emissiontype{II} IRT emission lines are saturated. 

In this study, we examined the infrared Mg\,\emissiontype{I} emission line at $\lambda 8807 \, \mathrm{\AA}$ (3d--3p transition) as a representative unsaturated chromospheric emission line. This Mg\,\emissiontype{I} emission line was isolated from other strong absorption lines.  It was detected with numerous chromospheric emission lines during a total solar eclipse in 1962 \citep{d68}. The solar imaging with the infrared Mg\,\emissiontype{I} emission line suggested that the emission line is formed in the chromosphere $500 \, \mathrm{km}$ above the photosphere \citep{f94}. As other example, the prominent infrared Mg\,\emissiontype{I} emission line was detected from eight T Tauri stars. %The objects showing saturated Ca\,\emissiontype{II} IRT emission lines may have a high rotation rate. 

Specifically, we investigated the infrared Mg\,\emissiontype{I} emission lines of 47 ZAMS stars using high-resolution spectral data. We examined the relationship between the Rossby number and the Mg\,\emissiontype{I} emission line strength. In the next section, we describe the observations and the data reduction procedure. In Section \ref{result}, we present the results, and in Section \ref{discussion}, we discuss the origin of the Mg\,\emissiontype{I} emission line and the correlation of the line strength and the Rossby number.

\section{Observations and Data Reduction}
\label{observ}

\subsection{Stellar parameters}
Our targets are F, K, and G-type ZAMS stars in IC 2391 ($50 \pm 5 \, \mathrm{Myr}$; \cite{ba04}) and IC 2602 ($30 \pm 5 \, \mathrm{Myr}$; \cite{st97}). The metallicity of both clusters has been determined to be close to the Sun, which is ${\rm [Fe/H]} = -0.01\pm 0.02$ for IC 2391 \citep{d09} and ${\rm [Fe/H]} = 0.00 \pm 0.01$ for IC 2602 \citep{ra01}. 

	A total of 52 ZAMS stars are confirmed as single stars in the IC 2391 and IC 2602 members based on the strength of the lithium $6708\, \mathrm{\AA}$ absorption line \citep{m09}. We examined the proper motion and radial velocity of these stars. The proper motion and the radial velocity are referred from {\it Gaia} Data Release 2 (DR2) \citep{g18}. For IC 2391, the mean proper motion in RA and Dec are $-23.6\pm4.7 \, \mathrm{km\cdot s^{-1}}$ and $22.0\pm4.4 \, \mathrm{km\cdot s^{-1}}$, respectively. For IC 2602, the corresponding values are $-17.3\pm1.0 \, \mathrm{km\cdot s^{-1}}$ and $10.8\pm1.3 \, \mathrm{km\cdot s^{-1}}$. Based on the observations in {\it Gaia} DR2 \citep{so18}, for 51 objects in IC 2391 and 325 objects in IC 2602, \citet{gu20} obtained mean radial velocities of $14.9 \pm 0.6 \, \mathrm{km \cdot s^{-1}}$ and $17.8 \pm 0.7 \, \mathrm{km \cdot s^{-1}}$, respectively. VXR PSPC 02A, VXR PSPC 31, VXR PSPC 78, and [RSP95] 96 have both proper motion and radial velocity that deviated by more than $3\sigma$ from the mean values. [RSP95] 42C has both proper motion and parallax that deviated by more than $3\sigma$ from the mean values. They were removed from our target list. A total of 47 targets were investigated in this study, which are presented in \tabref{tab:zams1}. 

\begin{small}
\renewcommand{\tabcolsep}{4pt}  
\begin{longtable}{lllclcllccc} 
\caption{Physical parameters of the ZAMS stars in IC 2391 and IC 2602.}
\label{tab:zams1}    
\hline\noalign{\vskip3pt} 
Object Name &   $r$  & $i$  &    $L/L_{\rm \odot}$ &  $T_{\rm eff}$ &  $(B-V)_0$ & dist & Period & $v \sin i$ &   $M_*$ & $\tau_{\rm c}$ \\  
{}  &  $\mathrm{[mag]}$ &  $\mathrm{[mag]}$ &   &   $\mathrm{[K]}$ & $\mathrm{[mag]}$ &  $\mathrm{[pc]}$ & $\mathrm{[days]}$ & $\mathrm{[km \cdot s^{-1}]}$ & $[M_{\odot}]$  & $\times 10^5 \, \mathrm{[s]}$\\ %&  $\mathrm{M_{\odot}}$ &  $\mathrm{Myr}$  %&  $M_{\rm conv}$ &  Age%$(B-V)_0$ &% $\mathrm{mag}$  & 
(1)  &  (2) &  (3) &  (4) &  (5) &  (6) & (7) &  (8) &  (9) & (10) & (11) \\    [2pt] % &  (11) 
\hline\noalign{\vskip3pt} 
%\hline
\endhead
\endfoot
\multicolumn{2}{@{}l@{}}{\hbox to0pt{\parbox{140mm}{\footnotesize 
{Reference of parameters. (2)(3) $r$- mag and $i$- mag: UCAC4 Catalogue \citep{za13}, ${}^a$ATLAS All-Sky Stellar Reference Catalogue \citep{t18} and ${}^b$\citet{m09}. }{(4)(5) Luminosity and $T_{\rm eff}$: {\it Gaia} DR2 \citep{g18} and ${}^b$\citet{m09}. }{(6) $(B-V)_0$: \citet{m09}. }{(7) Distance: {\it Gaia} DR2 \citep{ba18} and ${}^c$\citet{v07}. }{(8) Period: ${}^b$\citet{ps96} and ${}^b$\citet{bs99}. }{(9) $v \sin i$: \citet{m09}. }{(10) Canuto \& Mazzitelli [CM] Alexander model \citep{dm94}. }{(11) \citet{l10} model and \citet{noyes} model } %{(4) $B-V$: \citet{m09}. }
}\hss}} 
\endlastfoot
\multicolumn{6}{l}{IC 2391} & & & &  & \\ \hline
Cl* IC2391 L32  & 9.7 & 9.2          & 3.27     & 6590  &0.43  & 150        & -  & 68     & 1.5      &1 \\%      &                 &               \\
%VXR PSPC 02A    & 10.8        & 0.90 & 0.55     & 5015    & 150        & 0.23                & 235    & 1.0      \\%       & 0.1             & 20            \\
VXR PSPC 3A    &10.8  & 10.6       & 0.83     & 5607  & 0.67  & 150        & 3.93                & 10     & 1.0     &15 \\%        &                 &               \\
VXR PSPC 7     &9.5${}^a$   & 9.4       & 2.43     & 6472 & 0.45  & 146        & -  & 21     & 1.4  &2     \\%       &                 &               \\
VXR PSPC 12     &11.6     & 11.4    & 0.41     & 5042  &0.83  & 152        & 3.69             & 10     & 0.9    &25 \\%        &                 &               \\
VXR PSPC 14     &10.3     & 10.2    & 1.42     & 5768  & 0.56  & 158        & 1.35             & 43     & 1.1    &11 \\%        & 0.0             & 30            \\
VXR PSPC 16A   &11.5     & 11.3   & 0.44     & 5093  &0.87  & 152        & -                & 21     & 0.9   &25 \\%         & 0.1             & 30            \\
VXR PSPC 22A    &10.8  & 10.6     & 0.76     & 5587    & 0.73 & 148        & 2.31                & 8      & 1.0    &15 \\%        & 0.0             & 50            \\
%VXR PSPC 31     & 10.9        & 0.66 & 1.14     & 5716    & 203        & 3.03                & 17     & 1.0     \\%        & 0.0             & 100           \\
VXR PSPC 35A    &12.2    & 11.8     & 0.25     & 4982  &0.99  & 150        & 0.26             & 89     & 0.8  &25  \\%         &                 &               \\
VXR PSPC 44     &9.7    & 9.4     & 2.64     & 6536  & 0.41  & 151        & 0.57                & 79     & 1.4     &1 \\%       &                 &               \\
VXR PSPC 45A    &10.5${}^a$   & 9.8${}^b$     & 1.19     & 5125  & 0.80  & 151        & 0.22             & 235    & 1.2 &25  \\%          & 0.2             & 10            \\
VXR PSPC 50A    &-     & 11.6    & 0.21${}^b$     & 5210${}^b$  & 0.84  & 145${}^c$        & -                & 56     & 0.8   &20  \\%        & 0.1             & 100           \\
VXR PSPC 52     &10.2    & 10.1    & 1.48     & 5949  & 0.56  & 156        & 2.15 & 10     & 1.2    &11  \\%       &                 &               \\
VXR PSPC 62A    &11.6   & 11.3     & 0.45     & 4914  & 0.85  & 151        & 0.50${}^b$  & 49     & 0.9   &25 \\%         & 0.2             & 20            \\
VXR PSPC 66     &9.7    & 9.5     & 2.36     & 6382   &0.46 & 150        & 0.92  & 52     & 1.3      &2 \\%      & 0.0             & 20            \\
VXR PSPC 67A    &11.4   & 11.1    & 0.55     & 5004 &0.95   & 149        & 3.41  & 9      & 1.0    &25 \\%        & 0.4             & 10            \\
VXR PSPC 69A     &11.3   & 11.1   & 0.43     & 5062 &0.83   & 149        & 2.22  & 19     & 0.9   &25 \\%         & 0.0             & 50            \\
VXR PSPC 70     &10.6   & 10.4     & 0.99     & 5737 &0.63   & 152        & 2.61             & 16     & 1.0    &15 \\%        & 0.0             & 30            \\
VXR PSPC 72     &11.3  & 11.1     & 0.54     & 5336  &0.72  & 151        & 3.05             & 13     & 0.9    &20 \\%        & 0.1             & 30            \\
VXR PSPC 76A    &12.4  & 12.0     & 0.22     & 4776 &1.04   & 150        & 4.58             & 7      & 0.8   &25 \\%         & 0.2             & 30            \\
VXR PSPC 77A    &9.9   & 9.6      & 2.14     & 6163  &0.49  & 152        & 0.65             & 93     & 1.3    &7 \\%         &                 &               \\
%VXR PSPC 78     & 10.1        & 0.66 & 1.29     & 5786    & 145        & 1.28               & 46     & 1.1     \\%        & 0.0             & 20            \\
VXR PSPC 80A    &-   & 11.0${}^b$    & 0.38${}^b$     & 4880${}^b$ &0.96   & 145${}^c$        & -                & 145    & 0.9   &25   \\ \hline%       & 0.1             & 30            \\ \hline

\multicolumn{6}{l}{IC 2602}   & & & & & \\ \hline
Cl* IC2602 W79  &11.3     & 11.1   & 0.48${}^b$     & 5500${}^b$  &0.79  & 150        & 6.55                & 8      & 0.9      &31     \\
{[}RSP95{]} 1   &11.3    & 11.0   & 0.56     & 5198  & 0.87  & 149        & 3.85             & 7      & 0.9   &31  \\%        & 0.1             & 30            \\
{[}RSP95{]} 7  &9.4   & 9.0      & 3.74     & 6610    & 0.40& 150        & - & 58     & 1.5     &1 \\%       & 0.0             & 100           \\
{[}RSP95{]} 8A  &-      & 9.8${}^b$  & 1.27${}^b$     & 6190${}^b$  &0.61  & 145${}^c$        & -                & 27     & 1.2  &7  \\%         & 0.0             & 100           \\
{[}RSP95{]} 10  &-    & 11.6${}^b$    & 0.22${}^b$     & 4520${}^b$  &1.19  & 145${}^c$        & 3.16              & 14     & 0.8   & 42  \\%        & 0.2             & 30            \\
{[}RSP95{]} 14  &11.4   & 11.2     & 0.45     & 5071 &0.83   & 149        & 2.73               & 11     & 0.9   &31  \\%        & 0.0             & 50            \\
{[}RSP95{]} 15A &-   & 10.7${}^b$    & 0.48${}^b$     & 4920${}^b$ &0.89   & 145${}^c$        & 3.6${}^b$             & 7      & 1.0    &31  \\%       & 0.4             & 10            \\
{[}RSP95{]} 29  &12.2   & 11.9    & 0.23     & 4779 &1.07   & 150        & 2.22             & 21     & 0.8     &42  \\%      & 0.1             & 50            \\
{[}RSP95{]} 35  &10.4${}^a$  & 9.9${}^b$     & 1.16     & 5838  & 0.36  & 152        & 2.46             & 21     & 1.1     &11 \\%       & 0.0             & 100           \\
%{[}RSP95{]} 42C & 11.6${}^a$        & 0.76 & 21.84    & 6207    & 145${}^c$        & -                & 10     & 0.9             & 0.1             & 30            \\
{[}RSP95{]} 43  &11.8${}^a$  & 11.1${}^b$    & 0.37     & 4994 &0.91   & 150        & 0.78             & 47     & 0.9    &31  \\%       & 0.2             & 20            \\
{[}RSP95{]} 45A & - & 10.1${}^b$   & 0.97${}^b$     & 5960${}^b$ & 0.62   & 145${}^c$        & -                & 14     & 1.1     &11 \\%       & 0.0             & 100           \\
{[}RSP95{]} 52  &12.0   & 11.6      & 0.25     & 4828 &1.03   & 136        & 0.39             & 122    & 0.8   &31  \\%        & 0.2             & 30            \\
{[}RSP95{]} 58  &10.4   & 10.2      & 1.37     & 5502 &0.61   & 160        & 0.57             & 92     & 1.1    &18  \\%       & 0.0             & 30            \\
{[}RSP95{]} 59  &11.6   & 11.3      & 0.44     & 4961 &0.78   & 151        & 1.31             & 31     & 0.9    &31  \\%       & 0.1             & 30            \\
{[}RSP95{]} 66  &10.8   & 10.6      & 0.77     & 5476 &0.64   & 150        & 3.28             & 11     & 1.0   &18   \\%       & 0.1             & 20            \\
{[}RSP95{]} 68  &10.9   & 10.6      & 0.58     & 4908 &0.82   & 135        & 0.99             & 51     & 1.1  &31    \\%       & 0.5             & 7             \\
{[}RSP95{]} 70  &10.7   & 10.6     & 0.83     & 5755  &0.65  & 148        & 4.25             & 9      & 1.1    &11  \\%       & 0.0             & 100           \\
{[}RSP95{]} 72  &10.6   & 10.5      & 0.99     & 5557 &0.60   & 156        & 1.05             & 49     & 1.1  &18    \\%       & 0.0             & 100           \\
{[}RSP95{]} 79  &9.0${}^a$   & 8.6${}^b$      & 4.26     & 6383  &0.40  & 155        & 0.75         & 78     & 1.5  &1   \\%        & 0.0             & 100           \\
{[}RSP95{]} 80  &10.3   & 10.1      & 1.27     & 4980 &0.89   & 145        & 7.25       & 13     & 1.3   &31  \\%        & 0.5             & 5             \\
{[}RSP95{]} 83  &10.5   & 10.5      & 1.02     & 5588 &0.58   & 149        & 1.74                & 30     & 1.0    &18  \\%       & 0.0             & 30            \\
{[}RSP95{]} 85  &9.7    & 9.7      & 2.11     & 6171  &0.48  & 151        & 1.33           & 45     & 1.3    &7  \\%       & 0.0             & 100           \\
{[}RSP95{]} 88A &-     & 11.4${}^b$   & 0.25${}^b$     & 4350${}^b$  &1.16  & 145${}^c$        & 0.204${}^b$                & 255    & 0.8   &65   \\%       & 0.3  & 20            \\
{[}RSP95{]} 89  &12.5   & 12.0      & 0.19     & 4510  &1.20  & 146        & 4.73               & 9      & 0.8      &42  \\%     & 0.2             & 30            \\
{[}RSP95{]} 92  &10.0   & 9.9     & 0.65     & 5740   &0.63  & 145${}^c$        & 1.93               & 16     & 1.1  &11    \\%       & 0.0             & 20            \\
{[}RSP95{]} 95A &12.3  & 11.1      & 0.55     & 5071 &0.83  & 153        & 1.22                & 14     & 0.9      &31 \\   \hline \\ %      & 0.1             & 30            \\
%{[}RSP95{]} 96  & 11.9        & 1.21 & 0.23     & 4429    & 154        & 1.84                & 15     & 0.8      \\%       & 0.3             & 20            \\
\end{longtable} 
\renewcommand{\tabcolsep}{6pt}
\end{small}

\subsection{Data Reduction}
\subsubsection{Photometry}
To obtain the rotational period, $P$, we analyzed the light curves of 39 ZAMS stars. The members of IC 2391 and IC 2602 were observed in Transiting Exoplanet Survey Satellite ({\it TESS}) Sectors 8 and 10, respectively. All 39 stars were observed in the long-cadence ($1800$-second exposure) mode, which continued for $27\, \mathrm{days}$. These data were retrieved from the Multimission Archive at the Space Telescope Science Institute. We conducted principal component analysis using eleanor, an open-source tool for extracting light curves from {\it TESS} full-frame images \citep{f19}. After the photometric measurements, we searched for periodic signals by conducting Lomb--Scargle \citep{s82} periodogram analysis. For each object, the period of the light curve was determined. VXR PSPC 44 presented beating signatures in its light curve; therefore, we adopted the second maximum of the power. We calculated the amplitudes of the light curves by taking the 90th percentile flux and subtracting it from the 10th percentile flux. For three objects having ${\rm amplitude/mean \, flux \, error} < 10$ (Cl* IC2391 L32, VXR PSPC 7, and {[}RSP95{]} 7), we referred the periods listed in \citet{m09}. We used the periods or $v \sin i$ listed in \citet{m09} for three objects not observed by the {\it TESS}. The derived periods are listed in \tabref{tab:zams1}. They show good correlation with the periods measured by \citet{ps96} and \citet{bs99}, with a correlation coefficient of $0.997$. %The period of other 5 objects are not determined. 

\subsubsection{Spectroscopy}
We investigated the Mg\, \emissiontype{I} line, $\rm H\alpha$ line, and Ca\, \emissiontype{II} IRT lines of the considered 47 ZAMS stars using the $3.9 \, \mathrm{m}$ Anglo-Australian Telescope (AAT) and the University College London Echelle Spectrograph (UCLES) archived spectra. The principal investigator was S. C. Marsden. The dates of the observations were March 17--19, 2000; January 6--8, 2001; and February 11--12, 2001. The wavelength coverage was between $3522 \, \mathrm{\AA}$ and $9386 \, \mathrm{\AA}$ ($R \equiv \frac{\lambda}{\Delta \lambda} \sim 40,000 - 120,000$). The integration time for each object was between $300 \, \mathrm{s}$ and $1200 \, \mathrm{s}$. The number of frames obtained for each ZAMS star ranged from two to six. A detailed description of the observations is presented in \citet{m09}. 

We used the Image Reduction and Analysis Facility (IRAF) software package\footnote{IRAF software is distributed by the National Optical Astronomy Observatories, which are operated by the Association of Universities for Research in Astronomy, Inc., under a cooperative agreement with the National Science Foundation.} for data reduction. Overscan subtraction, bias subtraction, flat fielding, removal of scattered light, removal of cosmic rays and OH airgrow, extraction of a spectrum, wavelength calibration using a Tr--Ar lamp, and continuum normalization were conducted for all UCLES spectra. After the wavelength calibration, only 12 orders (order $64-67, 75-76, 84-87,$ and $94-95$) were reduced. 

To improve the $S/N$ ratio, we combined and averaged multiple frames for each object. Moreover, the UCLES spectrum overlaps with the next order over several tens of Angstroms. The average ratio of the intensity to the next order was $1.8$, which was independent of the wavelength. We applied a weight of $1.8$ to each short spectrum. 

\begin{figure}[htb]
	\centering
	\includegraphics[width=16cm]{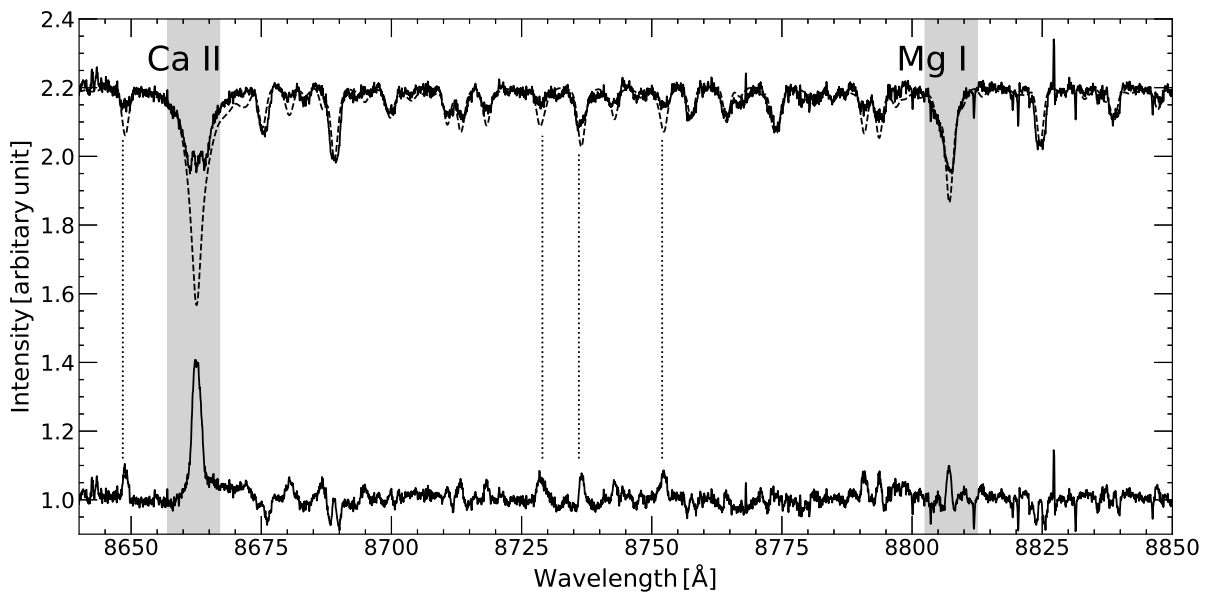} \caption{Procedure of spectral subtraction of photospheric component for ZAMS star. Observed spectrum of [RSP95] 43 is shown in top part of panel with solid line. Dotted line represents spectrum of fitted inactive template star. Spectra of [RSP95] 43 and template star are shown shifted by $+1.2$ for display purposes. Difference between spectra of [RSP95] 43 and template star is shown in bottom of panel, which is shifted by $+1$. Ca\,\emissiontype{II} line ($\lambda8662 \,\mathrm{\AA}$) and Mg\,\emissiontype{I} line ($\lambda8808 \,\mathrm{\AA}$) appear in emission. Other emission lines such as Si\,\emissiontype{I} ($\lambda8648, 8729, 8752 \,\mathrm{\AA}$) and Mg\,\emissiontype{I} ($\lambda8736 \,\mathrm{\AA}$) are also detected. } \label{fig:recx11}
\end{figure}

We subtracted the photospheric absorption component for all spectra. The Mg\, \emissiontype{I} emission component is typically buried by the photospheric absorption, as shown in \figref{fig:recx11}. Inactive stars with a spectral type similar to that of the target were used as template stars. We reduced the UCLES spectra of HD 16673 (F6V), $\rm \alpha$ Cen A (G2V), and $\rm \alpha$ Cen B (K1V). We also obtained the VLT archive spectra of ten inactive stars chosen from the inactive stars library \citep{yee} and the UVES POP library: $\zeta$ Ser (F2IV), HD 3861 (F8V), HD 1388 (G0V), HD 156846 (G1V), HD 109749 (G3V), HIP 94256 (G5V), HD 190360 (G7V), HD 217107 (G8V), HD 16160 (K3V), and HD 165341B (K4V). According to \citet{yee} and \citet{no04}, their metallicity is $-0.11 \leq {\rm [Fe/H]} \leq 0.31$. Their surface gravity is $4.01 \leq {\rm \log g} \leq 4.64 \, \mathrm{cm \cdot s^{-2}}$. The differences in the surface gravities of the target and template stars were sufficiently small, and therefore, were not expected to change the strengths of the photospheric absorption lines. For each template star, we reduced several frames, followed by averaging. The spectrum of an inactive star was shifted to match the radial velocity of the target star. For correcting the rotational broadening, the spectra of the template stars were convolved with a Gaussian kernel to match the widths of the absorption lines of each target. Using the high-$S/N$ spectra of the inactive stars, we carefully subtracted the photospheric absorption component. We considered the photospheric subtraction to be reliable if the photospheric absorption lines did not show any emission or absorption component. Many Fe\,\emissiontype{I} absorption lines appeared in the spectra of the targets and templates. However, the emission components of Fe\,\emissiontype{I} may be formed in a lower chromosphere \citep{v81}. Actually, T Tauri stars show narrow Fe\,\emissiontype{I} and Fe\,\emissiontype{II} emission lines at rest velocity, which are believed to be formed in the chromosphere \citep{hp92}. In contrast, we used Ni\,\emissiontype{I} ($6644, 6728, 7525\, \mathrm{\AA}$), Mn\,\emissiontype{I} ($6014, 6017, 6022\, \mathrm{\AA}$), V\,\emissiontype{I} ($6039\, \mathrm{\AA}$), and Ti\,\emissiontype{I} ($7523\, \mathrm{\AA}$) as the photospheric absorption lines. These absorption lines were isolated and relatively deep. We measured the standard deviation of the counts at Ni\,\emissiontype{I} ($\lambda 6644 \, \mathrm{\AA}$) after the subtraction of the spectrum of the inactive star, $\sigma_{\rm Ni\,\emissiontype{I}}$. The wavelength range was set as the wing width of the Ni\,\emissiontype{I} absorption line before the subtraction. $\sigma_{\rm cont}$ is the standard deviation of the counts between $\lambda 6683$ and $6693 \, \mathrm{\AA}$, where there was no strong feature. $\sigma_{\rm Ni\,\emissiontype{I}}/\sigma_{\rm cont} \sim 1$ represents the appropriate subtraction of the photospheric absorption. $\sigma_{\rm Ni\,\emissiontype{I}}/\sigma_{\rm cont}$ of the ZAMS stars were between $0.59$ and $2.51$. In principle, $\sigma_{\rm Ni\,\emissiontype{I}}$ is greater than $\sigma_{\rm cont}$, because the number of photons in the Ni\,\emissiontype{I} absorption line is smaller than that in the continuum. We considered that the photospheric subtraction was reliable if  $\sigma_{\rm Ni\,\emissiontype{I}}/\sigma_{\rm cont} < 1.2$. A total of 30 ZAMS stars showed $\sigma_{\rm Ni\,\emissiontype{I}}/\sigma_{\rm cont} < 1.2$, and the remaining 17 ZAMS stars showed $\sigma_{\rm Ni\,\emissiontype{I}}/\sigma_{\rm cont} \geq 1.2$. 

Before measuring the EQWs, the continuum components of the spectra were added to unity. To obtain the EQWs of the Mg\,\emissiontype{I} and ${\rm H\alpha}$ emission lines, the areas of the corresponding emission profiles were directly integrated. The EQW errors were estimated by multiplying the standard deviation of the continuum by the wavelength range of the emission lines of each ZAMS star. We estimated the standard deviations of the continuums near the Mg\,\emissiontype{I} and ${\rm H\alpha}$ emission lines. The wavelength range was $\lambda 8798-8802 \, \mathrm{\AA}$ and $\lambda 8813-8819 \, \mathrm{\AA}$ for the Mg\, \emissiontype{I} emission line, and that for the ${\rm H\alpha}$ emission line was $\lambda 6615-6623 \, \mathrm{\AA}$. In these wavelength ranges, no strong features were observed. We also measured the full width at half maximums (FWHMs) of the Mg\,\emissiontype{I} and Ca\,\emissiontype{II} IRT emission lines by fitting with a Gaussian function. The Ca\,\emissiontype{II} IRT lines at $\lambda 8542 \, \mathrm{\AA}$ of some template stars were out of range, and we did not measure their FWHMs. %These ranges are free from any emission and absorption lines. Finally, we measured the equivalent widths (${\rm EQW}$) of the Mg\, \emissiontype{I} emission line. %If no emission component was detected, we estimated an upper limit of equal width. We measured the standard deviation, $\sigma$, of the continuum around the center of the line. The wavelength range, $\Delta\lambda$, was set to be equal to FWHM of the lines of the template stars. Finally, we adopted $3\sigma \times \Delta\lambda$ as the upper limit of the equivalent width. 

\clearpage
\section{Results}
\label{result}

Figures 2 and 3 show the H${\rm \alpha}$, Ni\,\emissiontype{I}, and Mg\, \emissiontype{I} line spectra of the ZAMS stars after subtracting the photospheric absorption. Before subtracting the photospheric absorption component, the Mg\, \emissiontype{I} emission component is typically buried by the photospheric absorption, 11 ZAMS stars show H${\rm \alpha}$ as an emission line, and all ZAMS stars show the Ni\,\emissiontype{I} line as an absorption component. After the subtraction, the Mg\, \emissiontype{I} line is detected as an emission line in 45 ZAMS stars. The Mg\, \emissiontype{I} lines of these ZAMS stars show narrow emission, indicative of a chromospheric origin. Their EQWs range from $0.02 \, \mathrm{\AA}$ to $0.52 \, \mathrm{\AA}$. A total of 44 ZAMS stars show the H${\rm \alpha}$ emission line. The Ni\,\emissiontype{I} line does not present any emission component. The EQWs of the Mg\,\emissiontype{I} and H${\rm \alpha}$ emission lines and the FWHMs of the Mg\,\emissiontype{I}, H${\rm \alpha}$, and Ca\,\emissiontype{II} IRT emission lines are listed in \tabref{tab:discuss_of_obsZ}. 

Some of the chromospheric emission lines detected during a total solar eclipse \citep{d68} are also detected in the ZAMS spectra after the subtraction of the absorption components in the order of $64-67$ and $84-87$, P12($\lambda8750 \,\mathrm{\AA}$), P14($\lambda8598 \,\mathrm{\AA}$), He\,\emissiontype{I} ($\lambda6678 \,\mathrm{\AA}$), O\,\emissiontype{I} ($\lambda8446 \,\mathrm{\AA}$), Mg\,\emissiontype{I} ($\lambda8736 \,\mathrm{\AA}$), Si\,\emissiontype{I} ($\lambda6721, 8557, 8648, 8728, 8742, 8752 \,\mathrm{\AA}$), Ca\,\emissiontype{I} ($\lambda6573 \,\mathrm{\AA}$), and Fe\,\emissiontype{I} ($\lambda6518, 6575, 6634, 6713, 6753, 8680, 8686, 8713, 8790, 8824 \,\mathrm{\AA}$). Among them, P12, P14, He\,\emissiontype{I} ($\lambda6678 \,\mathrm{\AA}$), and O\,\emissiontype{I} ($\lambda8446 \,\mathrm{\AA}$) were also detected in the T Tauri star spectra obtained by \citet{hp92}. 

The emission line of Mg\,\emissiontype{I} at $\lambda 8807\,\mathrm{\AA}$ is the most frequently detected lines in the ZAMS spectra. The 45 ZAMS stars out of 47 show the emission lines of Mg\,\emissiontype{I} at $\lambda 8807\,\mathrm{\AA}$. The second most frequently detected line is the Mg\,\emissiontype{I} at $\lambda 8736\,\mathrm{\AA}$, which is detected in 40 ZAMS stars. However, this line blends with nearby Si\,\emissiontype{I} emission lines ($\lambda 8728, 8742 \,\mathrm{\AA}$). The Fe\,\emissiontype{I} ($\lambda 8689, 8790\,\mathrm{\AA}$) and Si\,\emissiontype{I} ($\lambda 8728 \,\mathrm{\AA}$) are detected in the 35 ZAMS stars, the 37 ZAMS stars, and the 30 ZAMS stars, respectively. Other chromospheric emission lines are detected in less numbers of the ZAMS stars, or blended with nearby chromospheric emission lines or OH airgrow. We thus concentrate the following discussion on the Mg\,\emissiontype{I} at $\lambda 8807\,\mathrm{\AA}$ and H${\rm \alpha}$ lines. %\footnote{Mg I線 ($\lambda 8807\,\mathrm{\AA}$)は47天体中45天体で輝線として検出された。次に多くの天体で検出されたのがMg I線 ($\lambda 8736\,\mathrm{\AA}$)で、40天体で輝線として検出された。それでも自転の速い天体では、近くの同等程度の強さの輝線とブレンドしてしまい、どこが輝線なのか分からなくなった。Fe I線 ($\lambda 8689, 8790\,\mathrm{\AA}$)はそれぞれ35天体と37天体で輝線として検出された。一方で、Si I線 ($\lambda 8728\,\mathrm{\AA}$)は30天体で輝線として検出された。他の線はもっと少ない天体において輝線として検出された。あるいは自転が速い天体では近くの彩層輝線やOH輝線とブレンドした。今回の天体のv sin iの最高値は$255\,\mathrm{km\cdot s^{-1}}$であり、ブレンドしないためには、他の線から$7.4\,\mathrm{\AA}$以上離れている必要がある。} 	%１．全ての天体で可視域のラインが輝線として出現するわけではない； He\,\emissiontype{I} ($\lambda6678 \,\mathrm{\AA}$),Si\,\emissiontype{I} ($\lambda6721 \,\mathrm{\AA}$), Ca\,\emissiontype{I} ($\lambda6573 \,\mathrm{\AA}$), and Fe\,\emissiontype{I} ($\lambda6518, 6575, 6634, 6713, 6753\,\mathrm{\AA}$) 	%２．今回の天体のv sin iの最高値は$255\,\mathrm{km\cdot s^{-1}}$である。ブレンドしないためには、他の線から$\,\mathrm{\AA}$以上離れている必要がある。高速自転星では他の彩層輝線やOH輝線と混ざるもの； P12($\lambda8750 \,\mathrm{\AA}$), N\,\emissiontype{I} ($\lambda8680, 8686 \,\mathrm{\AA}$), O\,\emissiontype{I} ($\lambda8446 \,\mathrm{\AA}$), Mg\,\emissiontype{I} ($\lambda8736 \,\mathrm{\AA}$), Si\,\emissiontype{I} ($\lambda8557, 8728, 8742, 8752 \,\mathrm{\AA}$), and Fe\,\emissiontype{I} ($\lambda8713, 8790, 8824 \,\mathrm{\AA}$)	%３．成長曲線法を参考に、Mg I輝線($\lambda8807 \,\mathrm{\AA}$)よりも$\log \, gf\lambda-\chi_m \theta$が小さいものを候補から外した。$\log \, gf\lambda-\chi_m \theta$, $g$は統計的重み、$f$は振動子強度、$\chi_m$はupper levelの励起ポテンシャルを示し、$\theta \equiv \frac{1\,\mathrm{eV}}{kT}\log_{10}e \sim \frac{5040\,\mathrm{K}}{T}$である。$\log\,gf$と$\chi_m$はKurucz et al. (1975)によるSemiempirical gf Valuesより引用した。$T$にはサハ・ボルツマンの式より推定した形成温度のピークを代入した(e.g. $4060\,\mathrm{K}$ for Mg I at $\lambda 8807\,\mathrm{\AA}$)。； O\,\emissiontype{I} ($\lambda8446 \,\mathrm{\AA}$), Mg\,\emissiontype{I} ($\lambda8736 \,\mathrm{\AA}$), Si\,\emissiontype{I} ($\lambda6721, 8557, 8648, 8728, 8742, 8752 \,\mathrm{\AA}$), Ca\,\emissiontype{I} ($\lambda6573 \,\mathrm{\AA}$), and Fe\,\emissiontype{I} ($\lambda6518, 6575, 6634, 6713, 6753, 8713, 8790 \,\mathrm{\AA}$)

\begin{figure}[htbp]
\begin{center}
    %\vspace{-2cm}
    \includegraphics[clip, width=17.5cm]{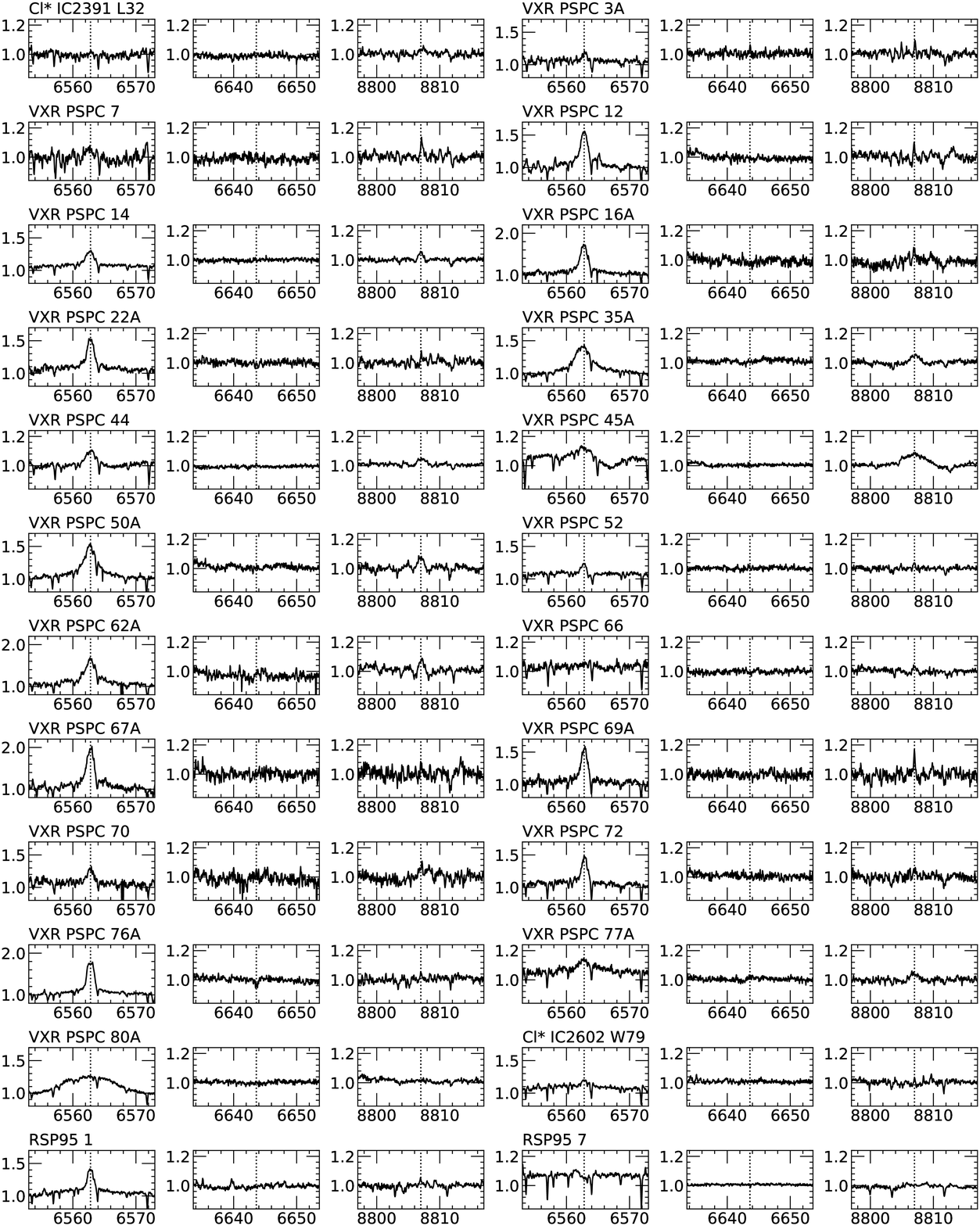}
    %\vspace{-4cm}
    \caption{H${\rm \alpha}$ ($6563\, \mathrm{\AA}$), Ni\,\emissiontype{I} ($6644\, \mathrm{\AA}$), and Mg\,\emissiontype{I} ($\lambda 8807 \, \mathrm{\AA}$) emission lines of ZAMS stars belonging to IC 2391 and IC 2602. Continuum is normalized to unity. Photospheric absorption lines are already subtracted.}
    \label{fig:result_emi1Z}
\end{center}
\end{figure}

\begin{figure}[htbp]
\begin{center}
    %\vspace{-2.5cm}
    \includegraphics[clip, width=17.5cm]{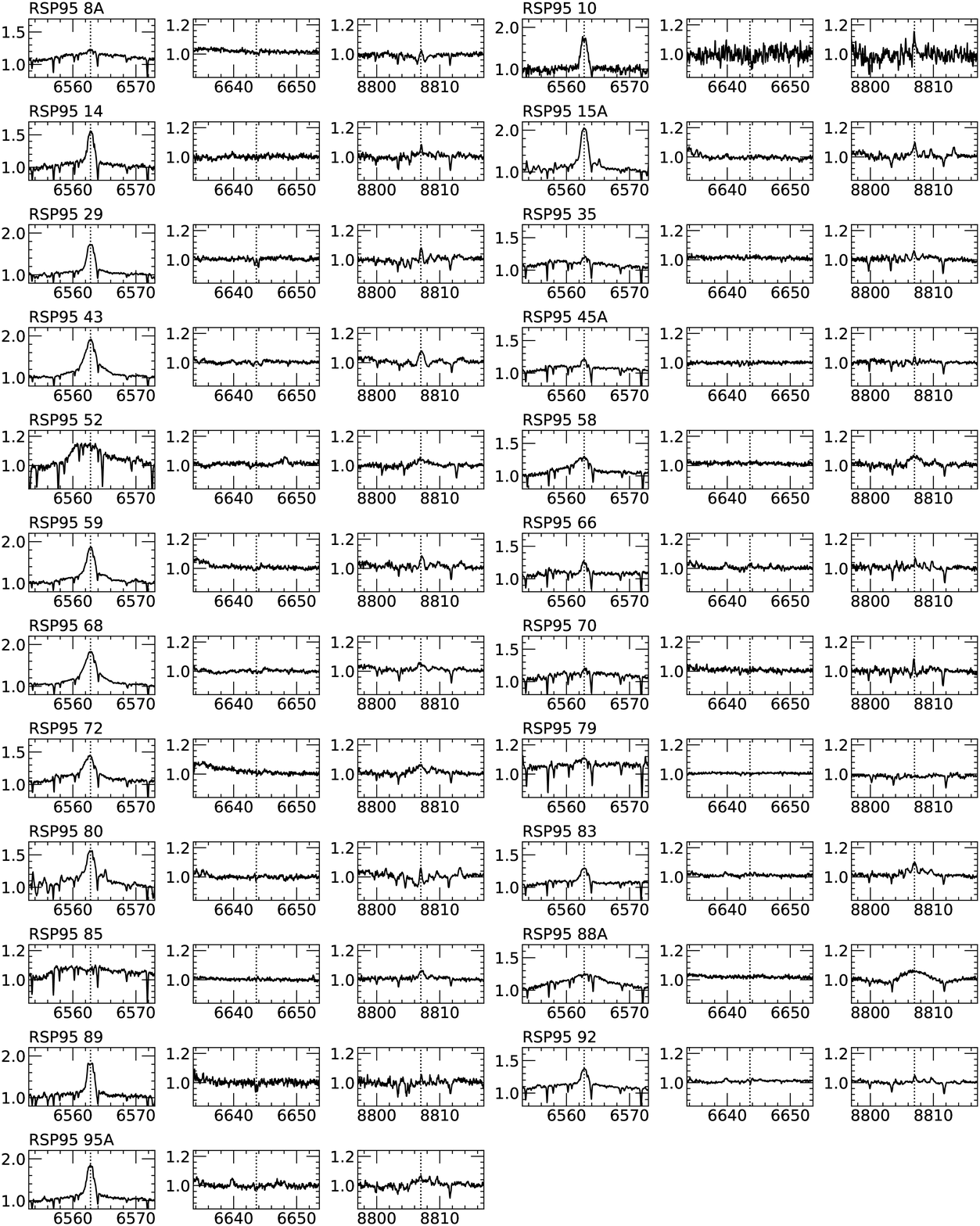}
    %\vspace{-3.5cm}
    \caption{H${\rm \alpha}$ ($6563\, \mathrm{\AA}$), Ni\,\emissiontype{I} ($6644\, \mathrm{\AA}$), and Mg\,\emissiontype{I} ($\lambda 8807 \, \mathrm{\AA}$) emission lines of ZAMS stars belonging to IC 2602. Continuum is normalized to unity. Photospheric absorption lines are already subtracted.}
    \label{fig:result_emi3Z}
\end{center}
\end{figure}
\section{Discussion}
\label{discussion}

\subsection{Mg\,\emissiontype{I} emission line}
\begin{figure}[ht]  \centering 
	\includegraphics[width=8cm]{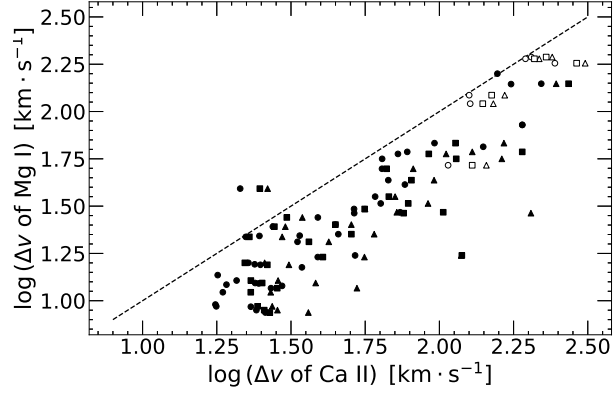} 
	\caption{Relationships between line widths $\Delta \nu$ of Ca\, \emissiontype{II} IRT emission lines and line width of Mg\,\emissiontype{I} emission line ($\lambda 8807 \, \mathrm{\AA}$). Filled symbols represent ZAMS stars in this study and open symbols represent T Tauri stars cited in \citet{hp92}. Circles, triangles, and squares symbols show Ca\, \emissiontype{II} IRT emission lines ($\lambda 8498, 8542, 8662 \, \mathrm{\AA}$), respectively. Dashed line represents $\Delta \nu \, {\rm of \, Ca\, \emissiontype{II} } = \Delta \nu \, {\rm of \, Mg\,\emissiontype{I} }$.} \label{fig:LWca}
	\end{figure}

	The line width of the Mg\,\emissiontype{I} emission line as a function of those of the Ca\, \emissiontype{II} IRT emission lines is plotted in \figref{fig:LWca}. We referred the measured line widths of six T Tauri stars from \citet{hp92}. These T Tauri stars show several prominent emission lines of Mg\,\emissiontype{I} and Mg\,\emissiontype{II}, which are considered to be formed in the chromosphere or the transition region. 

	The Mg\,\emissiontype{I} line width shows a positive correlation with each Ca\, \emissiontype{II} IRT line width. The Mg\,\emissiontype{I} emission lines in most ZAMS spectra are narrower than the Ca\, \emissiontype{II} IRT lines. This is consistent with \citet{hp92}, who claimed that these differences in the widths might be attributed to the different depths of the line formations, where the velocity fields are different. Based on the non-LTE solar model derived by \citet{v81}, the emission components of the Ca\, \emissiontype{II} IRT lines are formed $700--1200 \, \mathrm{km}$ above the photosphere. \citet{f94} suggested that the Mg\, \emissiontype{I} emission line is formed in the chromosphere, $500 \, \mathrm{km}$ above the photosphere. %Furmeister et al. (2005) also calculated the semi-empirical chromospheric model and revealed that Mg\,\emissiontype{I} emission line at $\lambda 3838 \,\mathrm{\AA}$ are formed at $\sim 4000\,\mathrm{K}$. Then we estimated the formation temperature of Mg\,\emissiontype{I} emission line at $\lambda 8807 \,\mathrm{\AA}$ with Saha equation and Boltzman equation. We obtained the ratio of the number densities of Mg\,\emissiontype{I} to Mg\,\emissiontype{II} with Saha equation, and the number densities of electrons at the upper level with Boltzman equation. We substituted the excitation energy, $7.6\,\mathrm{eV}$, and the potential energy of the upper level, $5.8\,\mathrm{eV}$, to Saha equation. The excitation energy and the potential energy of the upper level are refereed from the atomic database presented by NIST. As a result, the Mg\,\emissiontype{I} emission line at $\lambda8807 \,\mathrm{\AA}$ is most likely to be formed at $\sim 6000\,\mathrm{K}$. This calculation enabled us to know the Mg\,\emissiontype{I} emission line ($\lambda 3838 \,\mathrm{\AA}$) is similar to this, which potential energy of the upper level is $7.4\,\mathrm{eV}$. Moreover, $\log \, (gf\lambda) - \chi_m \theta$ of Mg\,\emissiontype{I} at $\lambda 3838 \,\mathrm{\AA}$ and $\lambda 8807 \,\mathrm{\AA}$ are $-13.5$ and $-13.1$, respectively. It is expected that the formation depth is approximately similar. According to the chromospheric model \citep{v81}, the temperature reaches at $6000\,\mathrm{K}$ at $\sim 1000\,\mathrm{km}$ above the photosphere. As same as this, the Ca \,\emissiontype{II} IRT emission line is most likely to form at $\sim 7000\,\mathrm{K}$, which corresponds to $2000\,\mathrm{km}$ above the photosphere. However, more accurate calculations will be needed such as non-LTE. %\footnote{日本語訳: サハの式よりMg IとMg IIの数密度の比を求め、ボルツマンの式より求めたupper levelにいる電子の数密度を掛け合わせた。NISTより引用したMg I輝線の励起ポテンシャル$7.6\,\mathrm{eV}$と、$\lambda 8807\,\mathrm{\AA}$のupper levelのエネルギーは$5.8\,\mathrm{eV}$、$\lambda 3838\,\mathrm{\AA}$のupper levelのエネルギーは$7.4\,\mathrm{eV}$を代入した。約$6000\,\mathrm{K}$の条件下でMg I輝線 at $\lambda 8807\,\mathrm{\AA}$は最も形成されやすく、$\lambda 3838\,\mathrm{\AA}$も同様であることが分かった。\citet{v81}による彩層の温度勾配モデルでは$6000\,\mathrm{K}$は光球の上空$1000\,\mathrm{km}$に相当する。同様の手法にて、Ca II三重輝線は約$7000\,\mathrm{K}$の条件下で最も形成されやすく、この温度は光球の上空$2000\,\mathrm{km}$に相当することが分かった。		ただし正確に計算する必要性もある。\citet{v73}ではCa II三重輝線は光球の上空$200-1300\,\mathrm{km}$で温度$4000-6000\,\mathrm{K}$の領域で、\citet{f05}のsemi-empirical model chromospheresではMg I at $\lambda 3838\,\mathrm{\AA}$で形成されると主張された。またMg I at $\lambda 3838, 8807\,\mathrm{\AA}$の$\log \, gf\lambda-\chi_m \theta$はそれぞれ$-13.5, -13.1$なので、おおよそ類似したformation depthであることが予想される。}%Actually, theoretical calculation have revealed that Ca \,\emissiontype{II} IRT emission line are expected to form $200--1300\,\mathrm{km}$ above the solar photosphere with $4000--6000\,\mathrm{K}$ \citep{v73}. 
	
Subsequently, we investigated the intensity of the Mg\, \emissiontype{I} emission line. To eliminate the dependence of the surface flux on the $T_{\rm eff}$ of the objects, we converted the EQW of the Mg\,\emissiontype{I} emission line into the ratio of the surface flux of the emission line to the stellar bolometric luminosity, $R^{\prime}_{\rm Mg\,\emissiontype{I}}$. This $R^{\prime}_{\rm Mg\,\emissiontype{I}}$ is similar to the parameter, $R^{\prime}_{\rm HK}$, derived from the Ca\,\emissiontype{II} H and K lines, as described by \citet{noyes}. In addition to Ca\,\emissiontype{II} H and K emission lines, $R^{\prime}_{\rm \lambda 8542}$, the ratio of the surface flux of the Ca\, \emissiontype{II} $\lambda 8542 \, \mathrm{\AA}$ line to the stellar bolometric luminosity, was previously used by \citet{so93} and \citet{jj97}. $R^{\prime}_{\rm \lambda 8498}, R^{\prime}_{\rm \lambda 8542}$, and $R^{\prime}_{\rm \lambda 8662}$ were used for the Ca\,\emissiontype{II} IRT emission lines by \citet{m09} and \citet{y20}. $R^{\prime}$ was used as a measure of the strengths of chromospheric emission lines, and its dependence on $N_{\rm R}$ has been examined. We first calculated the continuum flux per unit area at a stellar surface, $F$, as %. $F$ is given as
\begin{eqnarray}
	\log \frac{f}{f_0} & = & - \frac{2}{5} \times m_{i*},\\
	F & = & f \times \left( \frac{d}{R_*} \right)^2,
	\label{eq2}
\end{eqnarray}
where $f$ is the $i$-band continuum flux of the object per unit area as observed on the Earth. $m_{i*}$ is the apparent magnitude of the object in the $i$-band. The $i$-band continuum flux per unit area under $m_i = 0 \, \mathrm{mag}$ (the AB system) condition, $f_0$, is $1.852 \times 10^{-12} \, \mathrm{W \cdot m^{-2} \cdot \AA^{-1}}$ \citep{f96}. $d$ denotes the distance of an object from the Earth. $R_*$ is the stellar radius estimated using the Stefan--Boltzmann law with the photospheric luminosity and $T_{\rm eff}$ of the objects. Subsequently, $F$ was multiplied by the EQW of the Mg\,\emissiontype{I} emission line and converted into the surface flux, $F^{\prime} \, $, %, in each of Mg\,\emissiontype{I} emission lines; %\mathrm{W \cdot m^{-2}}
\begin{equation}
	F^{\prime} = F \times {\rm EQW}. 
	\label{eq3}
\end{equation}
Using $T_{\rm eff}$ of the target star, $R^{\prime}_{\rm Mg\,\emissiontype{I}}$ is calculated as % from \tabref{tab:zams1}
\begin{equation}
	R^{\prime}_{\rm Mg\,\emissiontype{I}} = \frac{F^{\prime} }{\sigma T_{\rm eff}^4}, 
	\label{eq4}
\end{equation}
where $\sigma$ is the Stefan--Boltzmann constant. The $R^{\prime}_{\rm Mg\,\emissiontype{I}}$ values for the ZAMS stars are listed in \tabref{tab:discuss_of_obsZ}. We also referred EQW of an Ca\, \emissiontype{II} IRT emission line ($\lambda 8542 \, \mathrm{\AA}$) from \citet{m09} and converted it into the ratio of the surface flux of the emission line to the stellar bolometric luminosity, $R^{\prime}_{\lambda 8542}$, using \equref{eq3} and \equref{eq4}.

\citet{noyes} and many authors used $N_{\rm R}$ as an indicator of stellar dynamo activity. 
\begin{equation}
\label{rossbym09}
N_{\rm R} \equiv \frac{P}{\tau_{\rm c}} = \frac{2 \pi R_*}{\tau_{\rm c} v \sin i} \langle \sin i \rangle. 
\end{equation}
We used $P$ for 42 ZAMS stars and $v \sin i$ for 5 ZAMS stars from \citet{m09}. Following the method of \citet{m09}, we assumed the random inclinations and multiplied the average value, $\langle \sin i \rangle = 0.785$, in \equref{rossbym09}. We estimated the $\tau_{\rm c}$ values of the ZAMS stars using the $\tau_{\rm c}$ models for the stars with the age of $50 \, \mathrm{Myr}$ (IC 2391) and $30 \, \mathrm{Myr}$ (IC 2602) presented in \citet{l10}. We fitted $\log \, \tau_{\rm c}$ as a third-order polynomial function of $\log T_{\rm eff}$. The $\tau_{\rm c}$ model is valid for objects with $T_{\rm eff} < 6810\, \mathrm{K}$. For six ZAMS stars with $T_{\rm eff} > 6180 \, \mathrm{K}$ (Cl* IC2391 L32, VXR PSPC 07, VXR PSPC 44, VXR PSPC 66, RSP95 7, and RSP95 79), we applied the approximation of $\tau_{\rm c}$ of \citet{noyes}, where $(B-V)_0$ is referred from \citet{m09}. %is out of range of the $\tau_{\rm c}$ model. For the six ZAMS stars

\begin{figure}[htb]
	\centering
	\includegraphics[width=8cm]{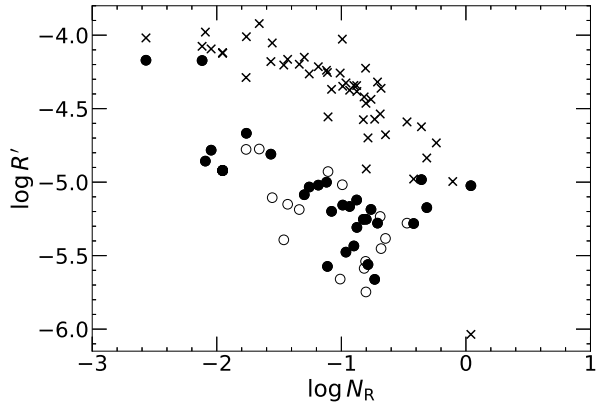}%210920/MgImt3_Nr_l10t.eps}%210803/MgImt3_Nr_l10.eps}
	\caption{Relationship between ratio of surface flux of Mg\,\emissiontype{I} emission line ($\lambda 8807 \, \mathrm{\AA}$) to stellar bolometric luminosity $R^{\prime}_{\rm Mg\,\emissiontype{I}}$ and Rossby number $N_{\rm R}$  of ZAMS stars (circles). Filled circles represent objects having $\sigma_{\rm Ni\,\emissiontype{I}}/\sigma_{\rm cont} < 1.2$. Open circles denote objects having $\sigma_{\rm Ni\,\emissiontype{I}}/\sigma_{\rm cont} \geq 1.2$. Cross symbols represent $R^{\prime}_{\rm \lambda 8542}$, which is ratio of surface flux of Ca\, \emissiontype{II} $\lambda 8542 \, \mathrm{\AA}$ line to stellar bolometric luminosity, of ZAMS stars.}\label{mgR2}% belonging to IC 2391 and IC 2602 
\end{figure}

The relationship between the ratio of the surface flux of the Mg\,\emissiontype{I} emission line to the stellar bolometric luminosity, $R^{\prime}_{\rm Mg\,\emissiontype{I}}$ and $N_{\rm R}$ of the ZAMS stars is shown in \figref{mgR2}. The $R^{\prime}_{\lambda 8542}$ values of the same ZAMS stars are also shown in the figure. When $N_{\rm R} > 10^{-1.1}$, $R^{\prime}_{\lambda 8542}$ increases with decreasing $N_{\rm R}$ until it saturates. When $N_{\rm R} < 10^{-1.1}$, $R^{\prime}_{\lambda 8542}$ reaches a constant level. Based on the discussion by \citet{m09}, chromospheric saturation is suggested for the objects with $N_{\rm R} < 10^{-1.1}$. %We referred $R^{\prime}_{\lambda 8542}$ from \citet{m09}. 

In contrast, $R^{\prime}_{\rm Mg\,\emissiontype{I}}$ increases with decreasing $N_{\rm R}$ even for the ZAMS stars with $N_{\rm R} < 10^{-1.1} $. The Mg\,\emissiontype{I} line is still unsaturated in the saturated regime for the Ca\, \emissiontype{II} IRT emission lines. The correlation coefficient between $R^{\prime}_{\rm Mg\,\emissiontype{I}}$ and $N_{\rm R}$ for the ZAMS stars showing $\sigma_{\rm Ni\,\emissiontype{I}}/\sigma_{\rm cont} < 1.2$ is $-0.70$. The derived linear fit is
\begin{equation}
	\log \, R^{\prime}_{\rm Mg\,\emissiontype{I}}  =  -0.3918 \log \, N_{\rm R} - 5.532.
\end{equation}
The correlation is valid even if other $\tau_{\rm c}$ models (\cite{noyes}, \cite{k96}, and \cite{g98}) are used. The Mg\, \emissiontype{I} emission line is considered to be a good indicator of chromospheric activity, particularly of active objects. 

VXR 45A and [RSP95] 88A show the largest $R^{\prime}_{\rm Mg\,\emissiontype{I}}$. Both are fast rotators; $v \sin i$ of VXR 45A and [RSP95] 88A are $235\,\mathrm{km\cdot s^{-1}}$ and $255\,\mathrm{km\cdot s^{-1}}$, respectively. FWHMs of the Mg\,\emissiontype{I} emission component correspond to $222\,\mathrm{km\cdot s^{-1}}$ and $159\,\mathrm{km\cdot s^{-1}}$. Both are comparable to the rotational velocity. It is known that the emission line is significantly broad if the line is formed by accretion \citep{y20}. Thus, we consider that the Mg\,\emissiontype{I} emission line is broad for these two stars because of their fast rotation. The two ZAMS stars also show broad emission components of Si\,\emissiontype{I} ($\lambda8728 \,\mathrm{\AA}$), but they are blended with other chromospheric emission lines.  %\footnote{日本語訳: VXR 45Aと[RPS 95] 88AのMg I線($\lambda 8807\,\mathrm{\AA}$)は、光球の吸収成分を除去する前は吸収線を示した。一方で質量降着を持つ前主系列星では、光球の吸収成分を除去しなくてもMg I線は輝線を示す(e.g. \cite{hp92})。そして$v \sin i$とFWHMに有意差があれば、質量降着による輝線であると言える。しかしVXR 45Aと[RPS 95] 88Aの$v \sin i$は$235 \,\mathrm{km \cdot s^{-1}}$と$255\,\mathrm{km \cdot s^{-1}}$なので、自転による広がりの可能性を棄却できない。一方で、VXR 45Aと[RPS 95] 88AのSi I線($\lambda 8728\,\mathrm{\AA}$)は、広い輝線を示しているようだが、どちらの天体も高速自転星なので他の彩層輝線とブレンドしてしまい、連続光成分とはっきりとした区別がつかない。}

Four objects located at Rossby number greater than $-0.5$ seem outliers, namely Cl* IC2391 L32, VXR PSPC 07, VXR PSPC 44, and VXR PSPC 66. This misalignment may be attributed to the difference of the applied $\tau_{\rm c}$ models. For the objects with $T_{\rm eff} \leq 6180\,\mathrm{K}$, we applied $\tau_{\rm c}$ model of \citet{l10}. However, the four outliers have $T_{\rm eff} > 6180\,\mathrm{K}$. $T_{\rm eff}$ higher than $6180\,\mathrm{K}$ is out of range of the model. Instead, we applied the approximation of $\tau_{\rm c}$ of \citet{noyes} for such high temperature objects. We constructed the relationship between $\tau_{\rm c}$ of Landin's model and $\tau_{\rm c}$ of Noyes' approximation for the ZAMS stars with $T_{\rm eff} \leq 6180\,\mathrm{K}$. The relationship is fitted by a linear function. We estimated the modified $\tau_{\rm c}$ for the four objects by extrapolating the linear function. The modified $\tau_{\rm c}$ is about $10^5 \,\mathrm{s}$ smaller than $\tau_{\rm c}$ estimated with Noyes' approximation, decresing $0.4$ in $\log\, N_{\rm R}$. As a result, the misalighment seems insignificant. %\footnote{日本語訳: これらの天体は負の相関から外れているようにも見える。この原因としては\citet{noyes}と\citet{l10}の$\tau_{\rm c}$のギャップであると思われる。この4天体はどれも$T_{\rm eff} > 6180 \,\mathrm{K}$であり、As a mentioned above, we applied the approximation of $\tau_{\rm c}$ of \citet{noyes}. $T_{\rm eff} < 6180 \,\mathrm{K}$の天体に対して$\tau_{\rm c}$ by \citet{l10} as a function of $\tau_{\rm c}$ by \citet{noyes}の回帰直線を得た。そして$T_{\rm eff} > 6180 \,\mathrm{K}$のZAMS 6天体について\citet{l10}の$\tau_{\rm c}$をextrapolateすると、どの天体もロスビー数が$-0.4$程度小さくなり、$\tau_{\rm c}$ by \citet{l10}を適用した天体に近づいた。	また、この4天体にはF6V std としてHD 16673のスペクトルを引いたという共通点があった。この標準星には問題ないと考えられる。まず$T_{\rm eff}$, $\log\,g$, ${\rm [Fe/H]}$などの物理量について他の文献も参照したが、特に不審な値は見当たらなかった。次に、他の標準星としてUVES POPより取り寄せたF6V型星HD 43318を試したが、$R^{\prime}$は$10^{-0.1}$程度しか変わらず、動きもランダムだった; 5天体中3天体の$R^{\prime}$が下がり、2天体の$R^{\prime}$が上がった。}%RSP95 7%This deviation can differ in the two types of applied $\tau_{\rm c}$ models. 

\citet{m09} argued that chromospheric saturation is caused by a similar mechanism to coronal saturation. \citet{s01} suggested that coronal saturation is a result of the coronal-emitting regions of a star being filled. \citet{m09} claimed that the emitting regions of Ca\,\emissiontype{II} of the ZAMS stars in $N_{\rm R} < 10^{-1.1}$ completely cover the chromosphere. The unsaturation of the Mg\,\emissiontype{I} emission line suggests that the area of its emitting region may be smaller than the Ca\,\emissiontype{II} emitting region area. 

The theoretical arguments on saturation are still being debated. The linear mean-field dynamo theory suggests that a small $N_{\rm R}$ allows exciting dynamo activity and leads to a strong magnetic strength. Therefore, the strengths of the Ca\,\emissiontype{II} emission lines should increase with decreasing $N_{\rm R}$. However, numerous observations evidence that the Ca\,\emissiontype{II} emission lines saturate with small $N_{\rm R}$. \citet{s13} explained the mechanism of the saturation based on Ohmic dissipation. They suggested that the ratio of the Ohmic dissipation rate to the power generated by the buoyancy increases with decreasing $N_{\rm R}$ for $10^{-2} < N_{\rm R} < 10^{-1}$. The ratio reaches maximum $0.8$ for $10^{-3} < N_{\rm R} < 10^{-2}$, and it is independent of the rotation. This implies that the dynamo activity is constant for a small-$N_{\rm R}$ object. %In $N_{\rm R} \lesssim 10^{-1}$, the ratio between Ohmic and total dissipation increases with decreasing $N_{\rm R}$. The ratio reaches at top, $0.8$, in $N_{\rm R} \sim 10^{-3} - 10^{-2}$. In an equilibrium state, the energy released by buoyancy in the models of \citet{s13} is dissipated by viscous dissipation and ohmic diffusion.\footnote{\citet{s13}; The ohmic diffusion requires that a magnetic field is built up by dynamo action and the rate of ohmic dissipation determines the fraction of the available power used for the magnetic field generation. For fast rotators fohm increases, this means that a larger fraction of the available power is converted to magnetic energy and dynamo action becomes more efficient.}\footnote{Christensen \& Aubert (2006); Dissipation and magnetic field strength are linked through the length scale of the field, or a dissipation timescale. Christensen \& Tilgner (2004) found  an inverse relation between the magnetic dissipation time $\tau^{\prime}$, that is, the ratio of magnetic energy to Ohmic dissipation, and the magnetic Reynolds number $R_{\rm m}$.} 

Many other absorption lines have been shown to also exhibit some filling in with chromospheric activity. \citet{t17} investigated in the difference between high- and low-activity spectra of $\rm \alpha$ Cen B. For the 48 nights, they generated 'relative' spectra by dividing each of the spectra by their lowest-activity spectra and obtained a large number of narrow emission lines such as Fe\,\emissiontype{I} $\lambda 4375, 4427, 4462\,\mathrm{\AA}$, Ti\,\emissiontype{II} $\lambda 4443.81\,\mathrm{\AA}$, and V\,\emissiontype{I} $\lambda 4444.21\,\mathrm{\AA}$. These features most likely originate from plage, spots or a combination of both. %They also found absorption components surrounding the Ti\,\emissiontype{II} emission peaks, which is considered to result from the difference in spot coverage. %\footnote{日本語訳: \citet{t17}は$\rm \alpha$ Cen Bの活動サイクルの調査として、48夜の分光観測を行った。最もactivityの低い日のスペクトルで、他の日のスペクトルを割った。Fe\,\emissiontype{I} $\lambda 4375, 4427, 4462\,\mathrm{\AA}$やTi\,\emissiontype{II} $\lambda 4443.81\,\mathrm{\AA}$、V\,\emissiontype{I} $\lambda 4444.21\,\mathrm{\AA}$も輝線として検出された。またTi\,\emissiontype{II}輝線の両脇に弱い吸収成分が残ったが、黒点の割合が日によって異なることが原因であると考えられた。}

It is probable that the intensity of the Mg\,\emissiontype{I} emission line has a positive correlation with the magnetic field strength. \citet{fl16} observed the Zeeman broadening of rapidly rotating G and K dwarfs, whose $N_{\rm R}$ was between $10^{-2}$ and $10^{0}$. It was showed that the large-scale magnetic field strength increases with decreasing $N_{\rm R}$. The magnetic field strength did not saturate with any Rossby number. \citet{c91} confirmed that the Mg\,\emissiontype{I} emission lines at $7 \, \mathrm{\mu m}$ and $12 \, \mathrm{\mu m}$ provide sensitive measures of the magnetic and electric field strengths of the solar atmosphere. The relation between the magnetic field strength and the Mg\,\emissiontype{I} emission line at $\lambda 8807\, \mathrm{\AA}$ needs to be investigated.

\subsection{ H${\rm \alpha}$ emission line} %Rotation-Activity Relation of 
We converted the EQW of the H${\rm \alpha}$ emission line into the ratio of the surface flux of the H${\rm \alpha}$ emission line to the stellar bolometric luminosity, $R^{\prime}_{\rm H{\rm \alpha}}$. We first calculated $F$ using \equref{eq2} and \equref{eq:rmag}.
\begin{equation}
	\label{eq:rmag}
	\log \frac{f}{f_0}  =  - \frac{2}{5} \times m_{r*}, 
\end{equation}
where $m_{r*}$ is the apparent magnitude of the object in the $r$-band. The $r$-band continuum flux per unit area under $m_r = 0 \, \mathrm{mag}$ (the AB system) condition, $f_0$, was $2.780 \times 10^{-12} \, \mathrm{W \cdot m^{-2} \cdot \AA^{-1}}$ \citep{f96}. Subsequently, $F$ was multiplied by the EQW of the H${\rm \alpha}$ emission line and converted into  $F^{\prime}$ using \equref{eq3}. By substituting $T_{\rm eff}$ and $F^{\prime}$ of the target star into \equref{eq4}, $R^{\prime}_{\rm H{\rm \alpha}}$ were calculated. The $R^{\prime}_{\rm H{\rm \alpha}}$ values for the ZAMS stars are listed in \tabref{tab:discuss_of_obsZ}. 

\begin{figure}[htb]
	\centering
	\includegraphics[width=8cm]{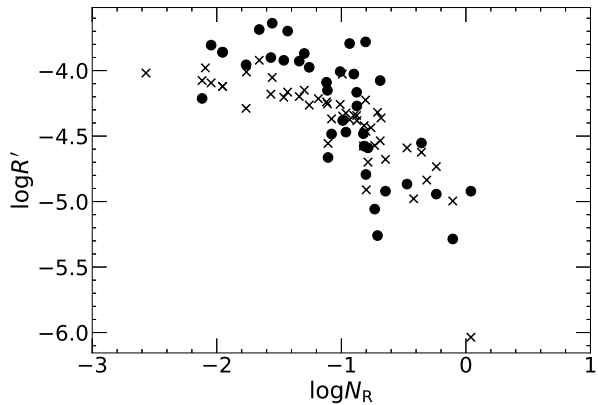}%210803/Halpha_l10.eps
	\caption{Relationship between ratio of surface flux of H${\rm \alpha}$ emission line ($\lambda 6563 \, \mathrm{\AA}$) to stellar bolometric luminosity $R^{\prime}_{\rm H \alpha}$ and Rossby number $N_{\rm R}$ of ZAMS stars (filled circles). Cross symbols represent $R^{\prime}_{\rm \lambda 8542}$, which is ratio of surface flux of Ca\, \emissiontype{II} $\lambda 8542 \, \mathrm{\AA}$ line to stellar bolometric luminosity, of ZAMS stars.} \label{fig:Halpha}
\end{figure}

The relationship between $R^{\prime}_{\rm H{\rm \alpha} }$ and $N_{\rm R}$ of the ZAMS stars is shown in \figref{fig:Halpha}. We also plot $R^{\prime}$ of the Ca\, \emissiontype{II} IRT emission line at $\lambda 8542\, \mathrm{\AA}$ referred from \citet{m09}. The ZAMS stars tend to show larger $R^{\prime}_{\rm H{\rm \alpha} }$ as $N_{\rm R}$ becomes smaller. In contrast, the ZAMS stars with $N_{\rm R} < 10^{-1.1}$ have $R^{\prime}_{\rm H{\rm \alpha} } \sim 10^{-3.9 \pm 0.5}$. This is the same as the results the Ca\, \emissiontype{II} IRT emission lines. However, the scatter of $R^{\prime}_{\rm H{\rm \alpha} }$ seems larger than that of $R^{\prime}_{\lambda 8542}$. It is likely that the extra scatter is due to the more strongly non-LTE nature of the H${\rm \alpha}$ source function \citep{cm79} and its strange optical depth behavior in cooler stars \citep{h95}. %This suggests that the scatter results from the telluric absorption lines blending with the H${\rm \alpha}$ emission line. 

\citet{ne17} investigated the relationship between $N_{\rm R}$ and $L_{\rm H \alpha}/L_{\rm bol}$, the H${\rm \alpha}$ emission line to stellar bolometric luminosity ratio for M dwarfs. The M dwarfs having small $N_{\rm R}$ showed larger $L_{\rm H \alpha}/L_{\rm bol}$ in $N_{\rm R} \geq 10^{-0.67}$. When $N_{\rm R} \leq 10^{-0.67}$, $L_{\rm H \alpha}/L_{\rm bol}$ reached a constant level at $1.85 \times 10^{-4}$. %Instead of subtracting the absorption components with any template spectrum, they estimated the strength of the absorption components numerically from the stellar mass. Then they correct the EQWs so that they can measure the amount of emission component above the maximim absorption level. \footnote{ただし\citet{ne17}は標準星を用いていないが、質量より吸収成分の等価幅を経験的に推定し、輝線の等価幅を算出した。} 

\citet{f17} observed the H${\rm \alpha}$, He\,\emissiontype{I}, Na\,\emissiontype{I}, and Ca\, \emissiontype{II} IRT lines of young stellar objects (YSOs) in the Lupus star forming region. One object did not show H${\rm \alpha}$ emission line; however, it was proved that the photospheric absorption component was filled with the chromospheric emission. Most of the young stellar objects show chromospheric emission lines with $R^{\prime}_{\rm H{\rm \alpha}} \leq 10^{-3.7}$. The ZAMS stars in our targets also show $R^{\prime}_{\rm H{\rm \alpha}} \leq 10^{-3.7}$. Before subtracting the absorption component, a portion of the objects showing featureless spectra at H$\alpha$ may indeed be chromospherically active stars. %\footnote{ここも修正点}

%\begin{tiny}%
\begin{small}
\renewcommand{\tabcolsep}{4pt}  
\begin{longtable}{p{24mm}p{16mm}p{16mm}p{8mm}p{8mm}p{8mm}p{8mm}p{8mm}p{9mm}p{12.5mm}p{9mm}p{9mm}}
\caption{EQWs, FWHMs, and $R^{\prime}$ of ZAMS stars belonging to IC 2391 and IC 2602. }
\label{tab:discuss_of_obsZ}
\hline
Object name           & \multicolumn{2}{c}{EQW $\, \mathrm{[\AA]}$}   & \multicolumn{5}{c}{FWHM $\, \mathrm{[km\cdot s^{-1}]}$}  & $\frac{\sigma_{\rm Ni\,\emissiontype{I}} }{\sigma_{\rm cont}}$  & $\log \, N_{\rm R}$ & \multicolumn{2}{c}{$\log \, R^{\prime}$} \\ \cline{4-8} \cline{11-12} %\cline{2-3} 
               & Mg\,\emissiontype{I}  & H${\rm \alpha}$  & Mg\,\emissiontype{I} & H${\rm \alpha}$ & Ca\,\emissiontype{II} & Ca\,\emissiontype{II} & Ca\,\emissiontype{II} &      &  & Mg\,\emissiontype{I} & H${\rm \alpha}$ \\
                              & $\lambda8807$  & $\lambda6563$  & $\lambda8807$  & $\lambda6563$ & $\lambda8498$ & $\lambda8542$ & $\lambda8662$ &      &  & $\lambda8807$  & $\lambda6563$ \\
\hline
\endhead
\endfoot
\multicolumn{2}{@{}l@{}}{\hbox to0pt{\parbox{170mm}{ {$^\dag 3 \sigma$ upper limit} }\hss}} 
\endlastfoot
\multicolumn{2}{l}{IC 2391}   &  &         &             &  &          &     &  &         &             &     \\ \hline
Cl* IC2391 L32 & 0.08 $\pm$ 0.07 & $<0.41 ^\dag$ & 61         &  -   & 78   & 129  & 76   & 0.88 & -0.31      & -5.17           &        -        \\
VXR PSPC 3A   & 0.05 $\pm$ 0.02 & 0.10 $\pm$ 0.05         & 9          & 38  & 18   & 27   & 17   & 1.22 & -0.65      & -5.38           & -4.92          \\
VXR PSPC 7    & 0.10 $\pm$ 0.03  & 0.10 $\pm$ 0.07         & 17         & 81  & 52   & 118  & 51   & 0.90  & 0.04      & -5.02           & -4.92          \\
VXR PSPC 12    & 0.04 $\pm$ 0.03 & 0.86 $\pm$ 0.09        & 9          & 60  & 24   & 28   & 24   & 0.71 & -0.90      & -5.43           & -4.03          \\
VXR PSPC 14    & 0.08 $\pm$ 0.02 & 0.34 $\pm$ 0.06        & 31         & 71  & 52   & 56   & 51   & 1.08 & -0.99         & -5.16           & -4.38          \\
VXR PSPC 16A   & 0.10 $\pm$ 0.04  & 0.93 $\pm$ 0.13        & 22         & 62  & 34   &   -   & 33   & 0.91 & -1.26      & -5.03           & -3.98          \\
VXR PSPC 22A   & 0.08 $\pm$ 0.03 & 0.57 $\pm$ 0.09        & 39         & 58  & 21   & 26   & 21   & 1.06 & -0.88      & -5.12           & -4.17          \\
VXR PSPC 35A   & 0.17 $\pm$ 0.07 & 1.53 $\pm$ 0.14        & 65         & 148 & 141  &   -   & 138  & 1.04 & -2.05      & -4.78           & -3.81          \\
VXR PSPC 44    & 0.12 $\pm$ 0.04 & 0.27 $\pm$ 0.05        & 56         & 79  & 64   & 162  & 63   & 0.97 & -0.36      & -4.98           & -4.55          \\
VXR PSPC 45A   & 0.50 $\pm$ 0.09  & 0.55 $\pm$ 0.10         & 140        & 222 & 174  &   -   & 171  & 0.92 & -2.12      & -4.17           & -4.21          \\
VXR PSPC 50A   & 0.16 $\pm$ 0.06 & 1.27 $\pm$ 0.12        & 41         & 117 & 77   &   -   & 75   & 1.14 & -1.76      & -4.67           &    -            \\
VXR PSPC 52    & 0.03 $\pm$ 0.01 & 0.21 $\pm$ 0.05        & 13         & 48  & 21   & 29   & 20   & 1.05 & -0.79      & -5.56           & -4.59          \\
VXR PSPC 62A   & 0.18 $\pm$ 0.05 & 1.07 $\pm$ 0.16        & 35         & 85  & 61   & 71   & 60   & 1.25 & -1.76      & -4.78           & -3.96          \\
VXR PSPC 66    & 0.06 $\pm$ 0.03 & $<0.39 ^\dag$ & 29         &  -   & 74   & 72   & 72   & 0.85 & -0.42      & -5.28           &       -         \\
VXR PSPC 67A   & 0.08 $\pm$ 0.06 & 1.65 $\pm$ 0.19        & 25         & 68  & 27   & 30   & 27   & 1.19 & -0.93      & -5.17           & -3.79          \\
VXR PSPC 69A   & 0.09 $\pm$ 0.03 & 0.59 $\pm$ 0.09        & 12         & 50  & 30   &  -    & 29   & 0.74 & -1.22      & -5.00              & -4.09          \\
VXR PSPC 70    & 0.06 $\pm$ 0.04 & 0.28 $\pm$ 0.10         & 28         & 56  & 39   & 35   & 38   & 0.59 & -0.82      & -5.25           & -4.48          \\
VXR PSPC 72    & 0.05 $\pm$ 0.03 & 0.47 $\pm$ 0.06        & 22         & 52  & 25   &   -   & 24   & 1.06 & -0.88      & -5.31           & -4.27          \\
VXR PSPC 76A   & 0.03 $\pm$ 0.02 & 1.60 $\pm$ 0.06         & 14         & 59  & 18   &   -   & 18   & 1.96 & -0.81      & -5.54           & -3.78          \\
VXR PSPC 77A   & 0.13 $\pm$ 0.05 & 0.21 $\pm$ 0.04        & 60         & 68  & 72   & 105  & 71   & 1.22 & -1.11      & -4.93           & -4.66          \\
VXR PSPC 80A   & 0.10 $\pm$ 0.10   & 2.22 $\pm$ 0.20         & 140        & 362 & 220  & 247  & 216  & 1.16 & -2.09      & -4.86           &     -          \\ \hline

\multicolumn{2}{l}{IC 2602}   &  &         &             &  &          &     &  &         &             &     \\ \hline
Cl* IC2602 W79 & 0.02 $\pm$ 0.01 & 0.08 $\pm$ 0.02        & 11         & 48  & 19   & 27   & 18   & 0.65 & -0.73      & -5.66           & -5.06          \\
RSP95 1        & 0.04 $\pm$ 0.02 & 0.34 $\pm$ 0.04        & 12         & 53  & 19   &   -   & 19   & 0.94 & -0.96      & -5.48           & -4.47          \\
RSP95 7        &  $<0.10 ^\dag$ & 0.06 $\pm$ 0.02 & - & 58  &   -   &   -   &   -   & 1.02 & -1.11      &    -             & -5.29          \\
RSP95 8A       & 0.06 $\pm$ 0.02 & 0.09 $\pm$ 0.19        & 20         & 582 & 33   & 42   & 33   & 1.10  & -0.76      & -5.19           &     -           \\
RSP95 10       & 0.07 $\pm$ 0.06 & 0.85 $\pm$ 0.17        & 12         & 52  & 25   &    -  & 24   & 1.16 & -1.18      & -5.02           &      -          \\
RSP95 14       & 0.03 $\pm$ 0.02 & 0.60 $\pm$ 0.06         & 9          & 51  & 23   &    -  & 23   & 0.71 & -1.11      & -5.57           & -4.15          \\
RSP95 15A      & 0.07 $\pm$ 0.04 & 1.81 $\pm$ 0.05        & 22         & 67  & 22   & 30   & 22   & 1.99 & -0.99      & -5.02           &    -            \\
RSP95 29       & 0.07 $\pm$ 0.02 & 1.09 $\pm$ 0.06        & 15         & 67  & 34   &   -   & 34   & 2.51 & -1.34      & -5.19           & -3.93          \\
RSP95 35       & 0.04 $\pm$ 0.02 & 0.05 $\pm$ 0.02        & 17         & 41  & 39   & 56   & 38   & 0.81 & -0.71       & -5.28           & -5.26          \\
RSP95 43       & 0.13 $\pm$ 0.03 & 2.01 $\pm$ 0.07        & 33         & 112 & 63   & 92   & 62   & 1.96 & -1.66      & -4.77           & -3.69          \\
RSP95 45A      & 0.03 $\pm$ 0.01 & 0.23 $\pm$ 0.02        & 12         & 42  & 24   & 38   & 23   & 1.29 & -0.68      & -5.45           &     -           \\
RSP95 52       & 0.13 $\pm$ 0.07 & 1.38 $\pm$ 0.10         & 85         & 217 & 190  &  -    & 187  & 1.07 & -1.95      & -4.92           & -3.86          \\
RSP95 58       & 0.17 $\pm$ 0.07 & 1.07 $\pm$ 0.09        & 68         & 163 & 96   & 165  & 94   & 0.97 & -1.57      & -4.81           & -3.90           \\
RSP95 59       & 0.08 $\pm$ 0.03 & 1.95 $\pm$ 0.07        & 22         & 84  & 46   & 60   & 45   & 1.35 & -1.43      & -5.15           & -3.70           \\
RSP95 66       & 0.02 $\pm$ 0.01 & 0.14 $\pm$ 0.03        & 9          & 41  & 26   & 36   & 25   & 1.25 & -0.80       & -5.75           & -4.79          \\
RSP95 68       & 0.08 $\pm$ 0.03 & 2.01 $\pm$ 0.06        & 43         & 110 & 67   & 96   & 66   & 1.47 & -1.55      & -5.11           & -3.64          \\
RSP95 70       & 0.06 $\pm$ 0.01 & 0.11 $\pm$ 0.02        & 12         & 44  & 27   & 53   & 27   & 1.64 & -0.47      & -5.28           & -4.87          \\
RSP95 72       & 0.09 $\pm$ 0.04 & 1.15 $\pm$ 0.08        & 50         & 111 & 64   & 82   & 63   & 1.19 & -1.30       & -5.09           & -3.87          \\
RSP95 79       & $<0.09 ^\dag$  & 0.10 $\pm$ 0.02 & -  & 92  & 60   &  -    & 59   & 1.10  & -0.24      &  -               & -4.94          \\
RSP95 80       & 0.07 $\pm$ 0.02 & 0.76 $\pm$ 0.06        & 16         & 77  & 23   & 26   & 22   & 1.34 & -0.69      & -5.23           & -4.08          \\
RSP95 83       & 0.08 $\pm$ 0.02 & 0.28 $\pm$ 0.03        & 25         & 52  & 45   & 51   & 44   & 1.20  & -1.08       & -5.20            & -4.48          \\
RSP95 85       & 0.07 $\pm$ 0.02 & $<0.20 ^\dag$ & 29         &   -  & 52   & 203  & 51   & 1.08 & -0.80      & -5.25           &      -          \\
RSP95 88A      & 0.52 $\pm$ 0.09 & 1.56 $\pm$ 0.18        & 159        & 456 & 157  &  -    & 154  & 0.78 & -2.57      & -4.17           &        -        \\
RSP95 89       & 0.02 $\pm$ 0.02 & 1.03 $\pm$ 0.08        & 10         & 56  & 18   &   -   & 17   & 1.80  & -1.01         & -5.66           & -4.01          \\
RSP95 92       & 0.03 $\pm$ 0.01 & 0.23 $\pm$ 0.02        & 15         & 50  & 25   & 31   & 24   & 1.86 & -0.82      & -5.59           & -4.58          \\
RSP95 95A      & 0.04 $\pm$ 0.02 & 1.06 $\pm$ 0.08        & 16         & 62  & 24   &  -    & 23   & 1.34 & -1.46      & -5.39           & -3.92          \\ \hline
\end{longtable}
\renewcommand{\tabcolsep}{6pt}
%\end{tiny} %}
\end{small} %}

\section{Conclusion}
\label{colclusion}

We investigated the infrared Mg\,\emissiontype{I} emission lines ($\lambda 8807 \, \mathrm{\AA}$) of 47 ZAMS stars belonging to IC 2391 and IC 2602. Archived data obtained using the AAT and the UCLES were used. After subtracting the spectra of the inactive stars from the ZAMS spectra, H${\rm \alpha}$, Ca\, \emissiontype{II}, and Mg\,\emissiontype{I} were detected as the emission lines. 

\begin{enumerate}
\item The infrared Mg\,\emissiontype{I} emission line width shows a positive correlation with each Ca\, \emissiontype{II} IRT line width. Most of the ZAMS stars show the narrower Mg\,\emissiontype{I} emission lines, instead of the Ca\, \emissiontype{II} IRT emission lines. The Mg\,\emissiontype{I} emission lines are detected from the most of the ZAMS stars. After subtracting the photospheric absorption components, the ZAMS stars with smaller Rossby numbers show stronger Mg\, \emissiontype{I} emission lines. The Mg\,\emissiontype{I} line is unsaturated even in ``the saturated regime for the Ca\,\emissiontype{II} emission lines,'' i.e., $N_{\rm R} < 10^{-1.1}$. The Mg\, \emissiontype{I} emission line is considered to be a good indicator of chromospheric activity, particularly for active objects. %\footnote{''This suggests that the Mg\,\emissiontype{I} emission lines are formed at different depths from the Ca\,\emissiontype{II} emission line-forming region.'' という文が入っていたが消す} 

\item The strength of the H${\rm \alpha}$ emission line is as strong as those of young stellar objects showing chromospheric emission lines. 
\end{enumerate}

 %\input{chapter5e}

%Ionization state the elements like ``Fe\,\emissiontype{II}'' can be expressed by 
% ``Fe\,\verb/\emissiontype{/II\verb/}/''. 

\begin{ack}
This study is based on the data acquired using the Anglo-Australian Telescope. We acknowledge the traditional custodians of the land on which the AAT stands, the Gamilaraay people, and pay our respects to the elders past and present. This research is also based on the observations made with the ESO Telescopes at the La Silla Paranal Observatory under programs ID 71.B-0529(A), 073.B-0607(A), 076.B-0055(A), 084.D-0965(A), 086.D-0062(A), 087.D-0010(A), 099.D-0410(A), and 266.D-5655(A). M. Y. was supported by a scholarship from the Japan Association of University Women, and would like to thank them. Y. I. is supported by JSPS KAKENHI grant number 17K05390. 
\end{ack}

%\begin{quote}
  
%\end{quote}

\end{document}